# Situation-Aware Integration and Transmission of Safety Information for Smart Railway Vehicles

Hanul Yeon and Dongsoo Har

Cho Chun Shik Graduate School for Green Transportation

Korea Advanced Institute of Science and Technology (KAIST)

Dajeon, Republic of Korea

{hanul.yeon, dshar}@kaist.ac.kr

**Abstract**

Recent trend of railway train development can be characterized in several aspects : high speed, infortainment, intelligence in driving, and so on. In particular, trend of high speed in driving is prominent and competition for high speed amongst several techno-savvy countries is becoming severe. To achieve high speed, engines or motors are distributed over multiple vehicles of train to provide increased motive power, while a single engine or motor has been mostly used for conventional trains. Increased speed and more complicated powertrain system naturally incur much higher chance of massive accidents. From this perspective, importance of proactive safety control before accident takes place cannot be over-emphasized. To implement proactive safety control requires situation-aware integration and transmission of safety information obtained from IoT sensors. Types of critical IoT sensors depend on situational conditions. Thus, integration and transmission of safety information should be performed with IoT sensors providing the safety information proper for faced situation. This brief paper is to devise a methodology how to operate IoT sensor network enabling proactive safety control for railway vehicles and to propose a queue management based medium access control scheme.

## 1. Introduction

Recent trend of railway train development can be characterized in several aspects : high speed, infortainment, intelligence in driving, and so on [1]. In particular, trend of high speed in driving is prominent and competition for high speed amongst several techno-savvy countries is becoming severe. To achieve high speed, engines or motors are distributed over multiple vehicles of train to provide increased motive power, while a single engine or motor has been mostly used for conventional trains [2]. Increased speed and more complicated powertrain system naturally incur much higher chance of massive accidents [3]. From this perspective, importance of proactive safety control before accident takes place cannot be over-emphasized. To implement proactive safety control requires situation-aware integration and transmission of safety information obtained from IoT sensors [4]. Types of critical IoT sensors depend on situational conditions. Thus, integration and transmission of safety information should be performed with IoT sensors providing the safety information proper for faced situation [5]. This paper is to devise a methodology how to operate IoT sensor network enabling proactive safety control for railway vehicles.

High speed train with engines or motors distributed over multiple vehicles are more subject to accidents such as unavoidable collision due to high speed, malfunction of complicated mechanics and electronics, derailing around the curved rail because of large centrifugal force. Therefore, it is important to proactively take steps not to face fatal situation. Critical safety information for avoiding fatal situation can be obtained from a single IoT sensor or a group of IoT sensors. Such sensors are placed in various spots of vehicles : in and around engine or motor, surface of vehicles, and inside vehicles [6]. Structure of connected sensors in a vehicle and across the vehicles forms an IoT sensor network. The structure of the IoT sensor network should be efficient in integration and transmission of collected IoT sensor data [7]. Type of main safety information obtained from the IoT sensor network varies according to faced situation. For instance, when train is driven in high speed, status of engine or motor, defect of wheels and axles, and detection of train position are particularly important. Thus, rate of collection and transmission of such information should be increased accordingly. Structure of IoT sensor network and applied channel access scheme should enable this type of flexible collection and transmission of safety information, regardless of channel state [8]. To this end, selecting critical sensors and transmitting their data should be smartly executed. As a whole, IoT sensor network with time-varying behaviour according to situations, e.g., speed and position of train, is desired to be implemented. Scope of this project is, therefore, designing IoT sensor network of which operation is adaptively changed to handle various situations.

IoT sensor network with fixed rate of data collection and data transmission is not capable of handling various situations. Suppose curved rail is ahead of running train. Train position detected by tag reader on the bottom of train indicates that curved segment of rail is imminent.

Figure 1 shows a derailing accident when the train did not proactively handle the operation around the curved segment of rail. In this situation, tilting sensor becomes critical for efficient operation of actuator counterbalancing the centrifugal force [9]. Since the centrifugal force is increased with higher speed of train, tilting sensor data and operation of actuator better be monitored more closely [10]. Defect of wheels and axles is also more critical with higher speed, so that rate of monitoring their defect should be adjusted according to speed. Thus, smart integration and transmission of IoT sensor data will enhance operational efficiency and provide proactive safety control.

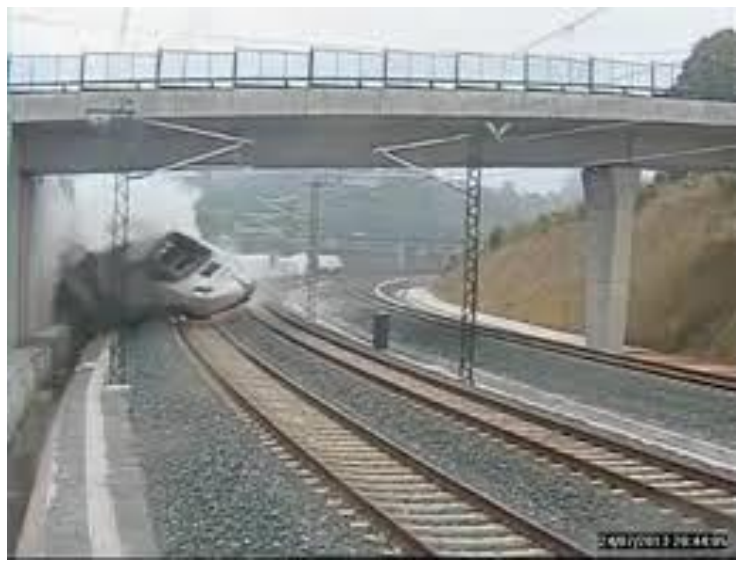

Fig.1 Tilting actuation or slowing down based on precise position detection of curved rail segment could help avoid derailing accident (courtesy of Google.com).

II. Solution for situation-aware integration and transmission of safety information
A. Collection of safety information

Approach to be attempted can be described twofold : situation-aware IoT sensor(+actuator) data processing and data transmission. When a train begins its operation, it needs to perform self-diagnosis to ensure initial safety control. When it is in operation (in driving), various types of IoT sensors for monitoring collect data to proactively perform safety control. Figure 2 show the reader placed on the bottom of a vehicle and the tag positioned along the track. Position of train is obtained by reader on bottom of vehicle(s) receiving position information from tag on track, or by other alternative means. If the immediate segment of rail from current position is curved, gyroscopic or other type of tilt sensor monitors the tilting condition more often and proper tilting actuation is executed when possible. In addition, as the speed of the train is rising, rate of data collected by video sensor for pantograph operation and wheel/axle defect sensors is to be increased. Figure 3 presents a defect detector for axle. On the other hand, interior IoT sensors installed in the interior of a vehicle, as seen in Fig.4, such as humidity sensor and fire sensor, for regular monitoring can take constant or close to constant data collection rates.

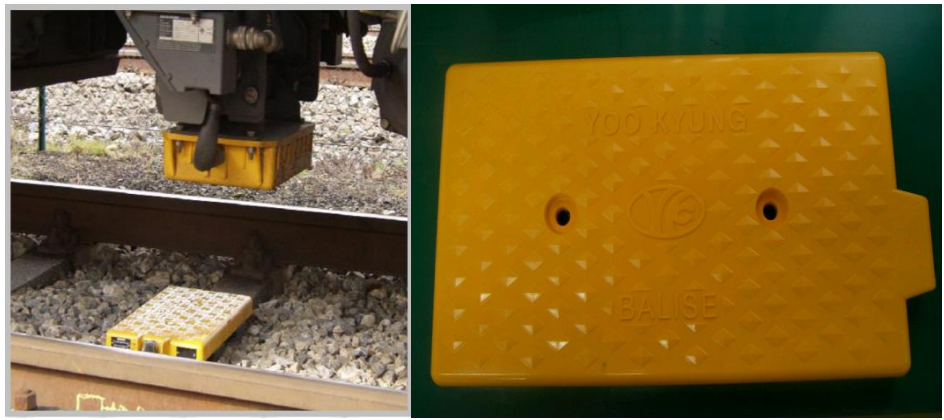

Fig. 2 Reader on vehicle and RFID tag on track for position detection

(courtesy of Korea Railroad Research Institute)

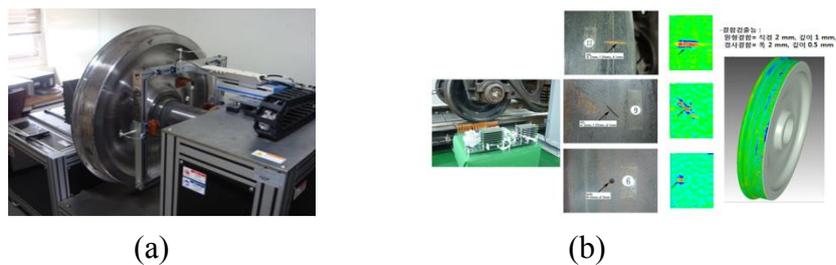

(a)        (b)

Fig. 3 (a) Axle defect sensor (b) Wheel defect sensor

(courtesy of Korea Railroad Research Institute)

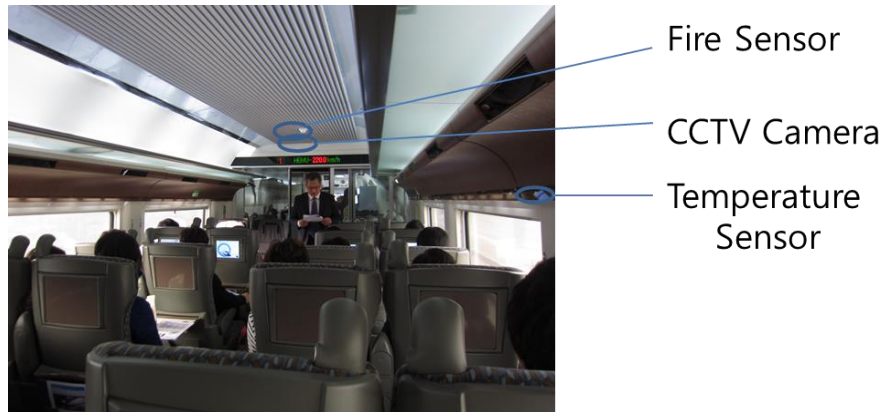

Fig. 4 Interior IoT sensors for regular monitoring

B. Situation-aware integration and transmission of safety information

Typically, operation of IoT sensors and actuators is determined by the situation alert transmitted by gateway of train or access points of vehicles. Important parameters transmitted to IoT sensors when alerting include speed, position, rail curvedness, and so on. Collection rate of IoT sensor data is dynamically changed for each IoT sensor, depending on type of faced situation (see Fig.5). IoT sensors such as humidity, pressure, acceleration, and so on can be wirelessly connected while others may be connected by wired links.

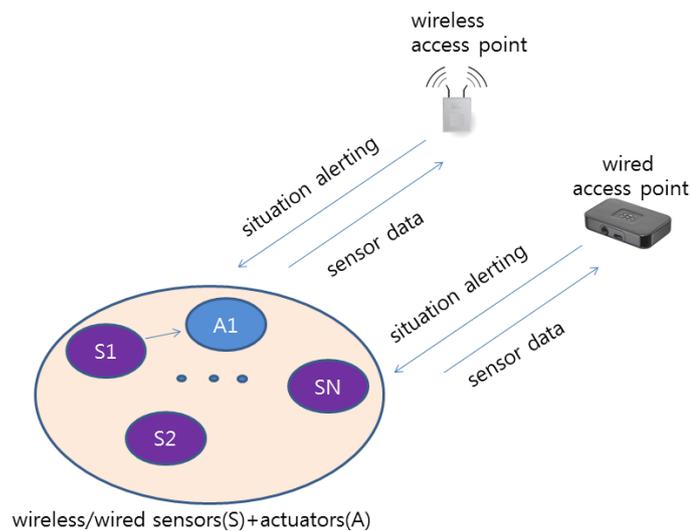

Fig. 5 Situation-aware operation of IoT sensor network. Collection rate of IoT sensor data is dynamically changed for each IoT sensor, depending on type of faced situation

Communication network within a vehicle can be formed in a combination of wired and wireless connections of IoT sensors. Some of the IoT sensor data are utilized for actuators without data transmission to external network and other data collected from the sensors are transmitted to external network. To deliver the IoT sensor data in an orderly manner, a single or multiple gateways of a train to outside communication networks, e.g., LTE-R cellular network, can be placed. Access to the gateway from each sensor can be direct or hierarchical.

C. Architecture for situation-aware integration and transmission

  C.1 Hierarchical architecture of sensor network

In a hierarchical structure as shown in Fig.6, cluster head which is a forwarding node to gateway can be considered as an intermediate node representing a vehicle. The sensor data are transmitted to the cluster head in each vehicle and forwarded to the gateway(sink) for eventual data aggregation. This architecture is beneficial when the number of sensors is large and coverage of whole sensor network is also large to some extent. Channel access for IoT sensor network can be prioritized according to data type and the priority of sensor data depends on situation even when they are obtained from the same sensor. Safety-critical IoT sensor data can be immediately transmitted to the external network by assigning a higher priority in contention or special time slots or even a separate gateway. This type of situation-aware data transmission scheme is also important. Various configurations and channel access schemes are tested to find out optimal solution.

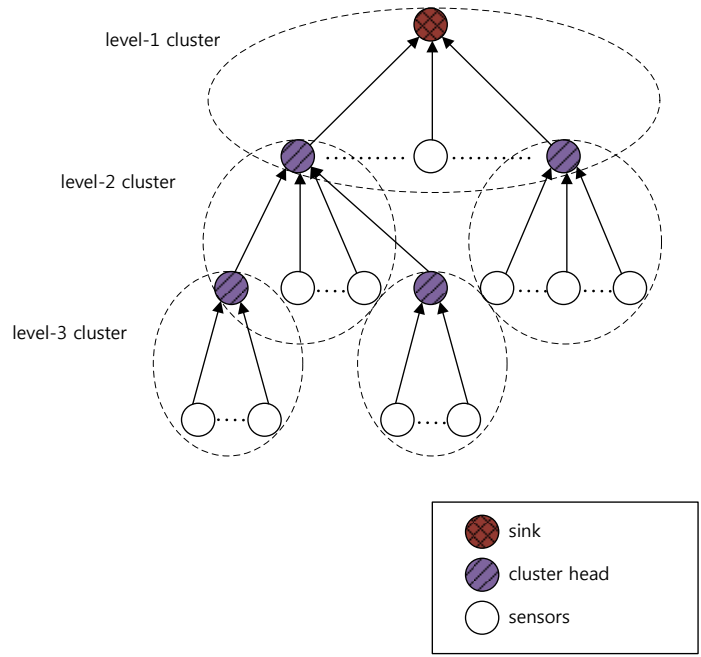

Fig. 6 Hierarchical IoT sensor network. IoT sensors can take time-varying priority in channel access to meet required low latency.

C.2 Queueing management based MAC for sensor network

Figure 7 queueing management for sensor network. The hierarchical network in Fig.6 is assumed operated in TDMA frame. Each cluster head assigns each group of sensors to a time slot. Sensors in a particular group has queue position. This queue is a virtual queue.

**Enqueue** – Coordinators can allocate new time-slot for new node according to the number of nodes in network, traffic, QoS needs of nodes, or add new node to previous time-slot which is not full of queue. New node added to previous time-slot can fulfill a transmission as soon as entering. Because queues is always added from the tail (n=N) to head (n=1) in backoff-queueing, queue which is not full always has space of slot 1. New node which fulfilled a transmitting enter the tail of queue where n=N.

Figure 7(b) shows the procedure where new nodes are enqueued to backoff-queue. Node A fulfills the transmitting as soon as (turn 1) entering the queue, and enters the tail of queue (n=4) in the next turn (turn 2). When the new node B enters the queue in turn 2, node B fulfills transmitting likewise. At the moment of turn 3 beginning, node A wins the slot 3 which lost 1 of n in turn 2, and node B wins the slot 4 which fulfilled the transmitting in turn 2. Because there is no entering of the new node in turn 3, node A fulfills the transmitting for its shortest backoff time. In the backoff-queue scheme, through this procedure, the queue operation is operated through the backoff of nodes in queue even though the coordinators do not manage queue.

**Deferring** – The nodes (deferring nodes) can oppose to fulfill transmission if they do not want transmission though the transmitting order came due to the queuing. In this case, the node which has longer backoff than deferring node gets to have an authority of transmitting. Deferring node has same queue position in the next turn, and the node which fulfilled the transmission enter the tail of queue in next turn.

**Dequeue** – In the backoff-queueing scheme, to prevent many nodes have same backoff slot, there should be no empty slot from the tail like shown in the process of enqueue. The process of enqueue should maintain the sorting state of queue with the certain steps in case that a node in time-slot is removed (excluded from network) though the process is operated without empty slot from the tail. If a node want to escape from time-slot (out node), should wait until its transmitting turn comes. The out node which attained the transmitting right deliver the intention of dequeue and queue-position of dequeue to coordinator. Coordinator receives this information from the out node and order queue sorting to node in that time-slot in next turn. Because queue sorting ordering of coordinator is fulfilled without backoff, it is fulfilled faster than transmitting of nodes in queue. The nodes which received the queue sorting order adjust their position based on the last position information of out node. If the last turn's position of out node which delivered its dequeuing intention to coordinator is bigger than the position of out node, the nodes apply the last turn's position to this turn. A node whose position is smaller than the position of out node gets to have position bigger than the last turn's position by 1. Therefore, the sorting of queue is finished.

Figure 7(c) shows the procedure of queue sorting according to the node dequeue. Backoff-queue is state of full. And node A, B, C, and D are waiting in the slot 1-4 respectively. In turn 1, the out node is B and if node B acquire the transmitting right from A's deferring transmission, it transmit the dequeuing intention and present queue position to coordinator. The coordinator transmits the dequeuing intention and position of node B to the nodes in queue in the next turn (turn 2). Node A whose position was smaller than the position of node B in turn 1 has the position increase by 2 than in turn 2, and other nodes maintain their position in turn 1 and sorting of queue.

**Garbage-slot collection** – In backoff-queue scheme, because the position of each node in time-slot is only known by itself, if a random node quitted the communication without dequeuing procedure (power-off, breakdown), other nodes cannot know that the nodes of queue decreased. Therefore, if the number of nodes which quit without dequeuing procedure increases, the network efficiency decreases. Likewise, unreturned backoff-slot is called garbage-slot. To collect the garbage-slot, the coordinator should regularly check if each node is alive, and if the coordinator recognizes the node as having been quitted if the reaction is continuously absent, it fulfills the collection to retrieve the garbage-slot. The collecting method is to make all the nodes in time-slot during the N turn which is length of queue in time-slot fulfill the transmitting so that the garbage-slot is removed and queue is resorted. Once the collection is begun, collection counter increases for each turn from 1 to N. Each node memorizes the position at the moment when the collection is begun, and fulfills the transmission in case of the collection counter is equal to the value. All the nodes examine whether the transmission is fulfilled through the carrier sensing during the collection, and decrease the position value by one in case of the transmission is fulfilled. The nodes which fulfilled the transmission during the collection enter the tail of queue. After the N turns, the queues are sorted, and the coordinator modify the number of nodes in time-slot based on the actual fulfillments of transmission in N turns.

Figure 7 (d) shows the process of garbage-slot collection. Backoff-queue is state of full, and node A, B, C, and D occupy the slot 1-4 respectively, but because node B is quitted without dequeuing procedure, slot 2 became garbage-slot. When the coordinator orders the collection. each node fulfills the transmission in order according to the collection counter.

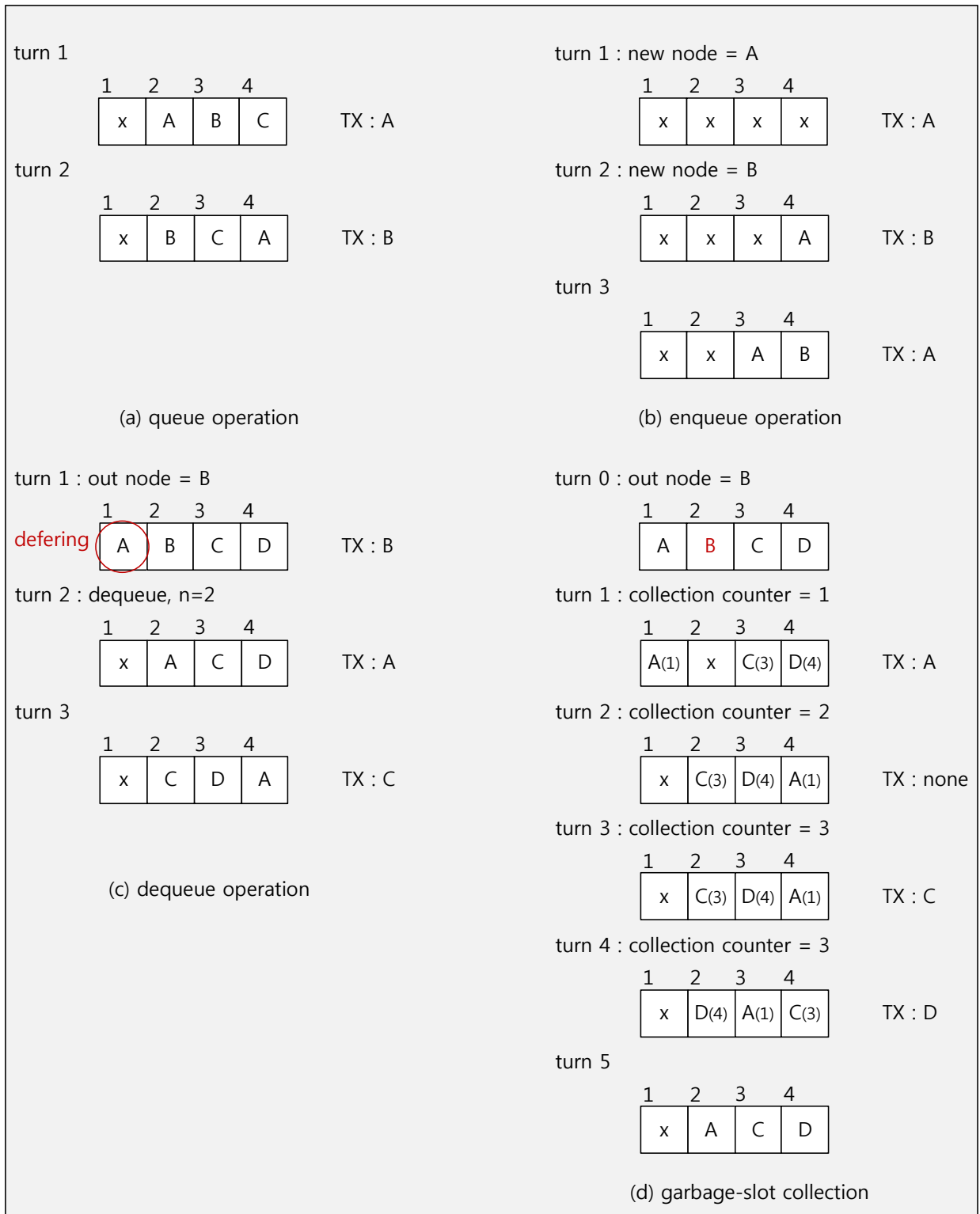

Fig. 7. Principal operations of backoff-queueing scheme.

With foregoing queue management and carrier sensing by each sensor, performance of the sensor network is obtained as follows.

**Low rate communication** – To evaluate the performance of low rate transmitting, the slotted beacon mode is realized which fulfills contention access using beacon and time-slot from the channel access scheme of 802.15.4. And the backoff-queueing proposed with this basis was applied. 15 time-slots were used in both original 802.15.4 and backoff-queueing, and the beacon interval was set to 122.88ms which is minimal value of 802.15.4. The backoff slot is 320us.

There are 60 nodes included in the network, and the nodes are divided into three groups according to the transmission interval. The transmission interval of group A is 100ms and that of group B is 250ms. The group C is composed of the nodes whose transmission interval is random between 100ms and 1s. The transmitting payload of each node is assumed to be equal to 32byte. Group A and B are the nodes fulfilling continuous monitoring, and group C is composed of the nodes fulfilling transmission as occasion. In the experiment, the ratio of nodes of group A, B, and C was 1:1:1, 1:1:2, or 1:1:4. In the dementia assistive system, the sensors deliver the data to hub and collect the data. Thus, this experiment is set where all the sensors transmit data to one network hub by uplink. The performance was evaluated by the standard of delay between the creation of node data of each group and the transmission fulfillment.

Figure 8 represents the graph of performance of 802.15.4, B-MAC, and [backoff-queueing] scheme in low rate communication. The number of nodes in the network is 60, and the horizontal axis represents the ratio of node group, and the vertical axis represents the mean delay time of each transmission.

In case that the ratio of group A, B, or C is 1:1:1, 1:1:2 or 1:1:4, backoff-queueing scheme shows higher performance than the original 802.15.4. When the ratio of group A, B, and C is 1:1:1, while the original 802 shows the delay of 112.0ms and B-MAC shows 109.2ms, the average delay in backoff-queueing was 49.0ms. When the ratio of group A, B and C was 1:1:2 or 1:1:4, the average delay of original 802.15.4 was 105.4ms, 106.5ms respectively, and B-MAC showed the average delay of 57.3ms, 83.9ms respectively. These results mean that the more number of nodes which transmit regularly, the less average delay. And in the circumstance of DAS whose sensor action is successive and periodic, the backoff-queueing method seems more appropriate than the original 802.15.4.

**High rate communication** – To measure the performance in high rate, the situation channel access is fulfilled based on contention with enhanced distributed channel access(EDCA) which is used in 802.11 standard was assumed, and the performance with backoff-queueing scheme is compared. The services such as VoIP, video phone, audio/video streaming used the high rate, and the offered load was 0.096Mbps, 0.5~1 Mbps, and 0.128-24 Mbps respectively.

The number of nodes is 20 at most. The ratio of VoIP, video phone, AV streaming node referred to the dementia assistive system circumstance where there are 1 VoIP, 1 video phone, and the remaining is video streaming for patient monitoring. The composition of nodes is star topology focused on WLAN access point, and it is assumed that there is no transmission between non-AP nodes.

The performance measurement is based on the averaged transmission delay, and the number of nodes was increased to 10, 15, and 20. The backoff time of backoff-queueing was set to be 20us, and the time-slot was set 15. And the back off time of 802.11ac was set 20us, and the minimum size of contention window was set 31, and the maximum size of that was set 1023. The channel capacity is 667MHz according to the modulation and coding scheme 9 of 802.11ac.

Figure 9 is a graph representing the performance of original 802.11 EDCA and backoff-queueing in the high rate communication. The horizontal axis represents the increase of the nodes' number, and the vertical axis represents the average delay of each transmission.

In the environment where the number of nodes was 10 or 15 which is smaller than or equal to time slot of backoff-queueing scheme, the delay of backoff-queueing scheme was 15.1ms or 56.2ms respectively, and the delay of original 802.11ac was 121.7 ms or 132.5 ms respectively. When the number of nodes was 20, the average delay of original 802.11 DCF was 139.3 ms, and it was improved to 118.2 ms when the backoff-queueing was applied.

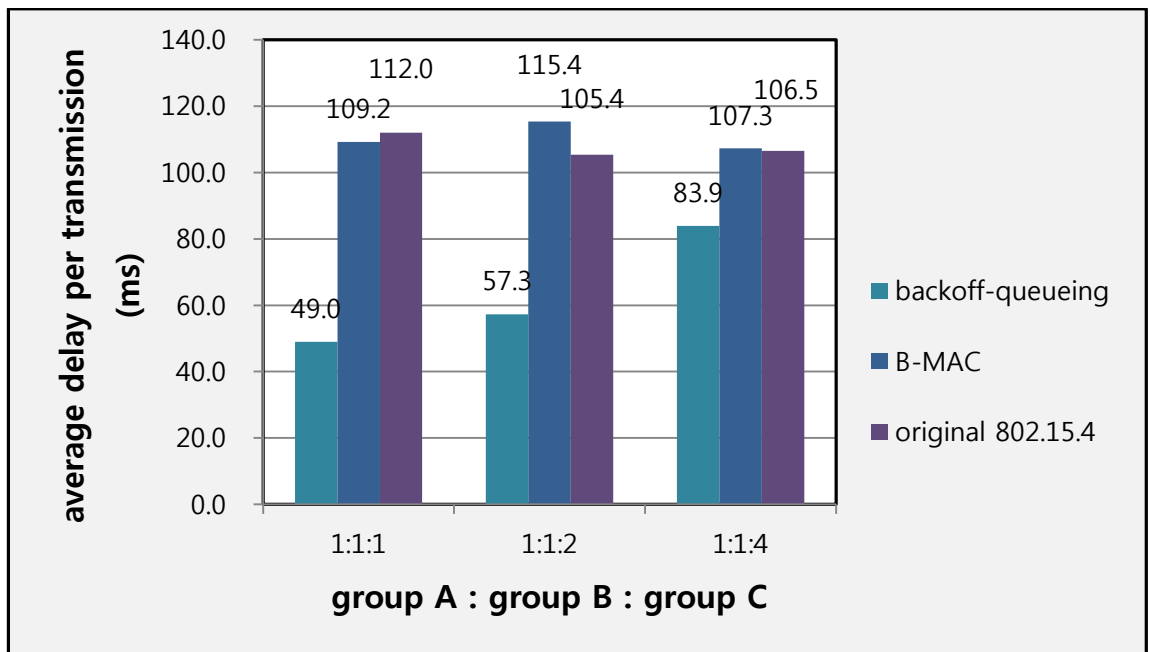

Figure 8. average delay per transmission of low rate network

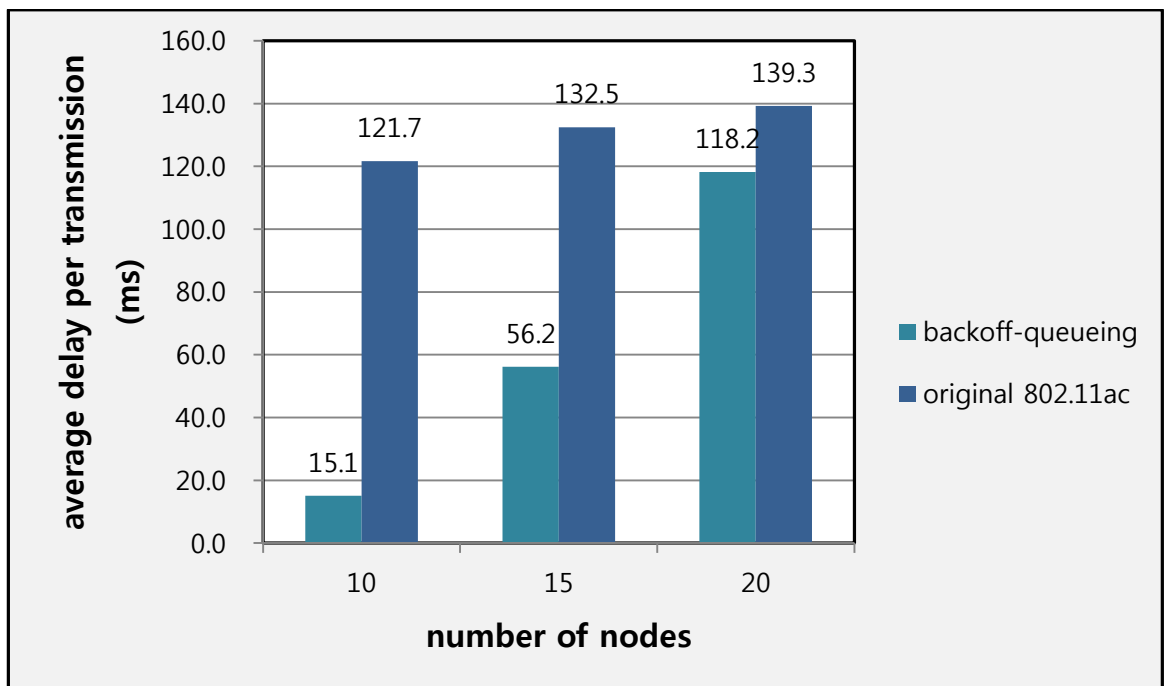

Figure 9. average per transmission of high rate network

D. Innovation

Innovation in situation-aware integration and transmission of safety information can be described in terms of effectively changing structure of IoT sensor network according to faced situation and unique manner of network operation. Situation-aware integration and transmission of safety information has several innovative technologies as follows

1) Adaptive IoT sensor data collection to efficiently handle safety-critical situations
2) Dynamic change of channel access priority of IoT sensors to meet requirements of safety-critical applications
3) Scheduling-based channel access scheme optimized for smart railway vehicles.
4) Effectively self-organizing IoT sensor network by time-varying weighting without hardware reconfiguration
5) Operational efficiency for wide range of train speed

Approach of situation-aware integration and transmission of safety information has several differentiating technologies. They can be depicted in terms of structure of IoT sensor network and adaptive operation of the IoT sensor network.

Table I. Differentiators of situation-aware collection and transmission of sensor data from current approaches.

|  | State-of-the-art IoT sensor network | Proposed IoT sensor network |
|---|---|---|
| **network structure** | fixed | dynamic by dynamic weighting |
| **network operation** | static | adaptive to different situations |
| **channel access priority** | data class or # of failure dependent | situation dependent |
| **train speed dependency** | independent in operation | speed dependent |
| **optimization in various situations** | No | Yes |
| **hardware implementation cost** | essentially the same ||

E. Competitiveness

Situation-aware approach for integration and transmission of IoT sensor data is essential to meet safety requirements of railway vehicles. Time-varying and adaptive behaviour of IoT sensor network might demonstrate disruptive nature of IoT sensor network for railway vehicles. Provision of critical safety information practically depends on smart operation of IoT sensor network. To do so requires designing of train-specific IoT sensor network. Train-specific IoT sensor network should be flexible in its operation to efficiently handle various situations during train service. Structure and operation of situation-aware IoT sensor network is a key contributor to development of next generation train. Since the methodology how to build situation-aware IoT sensor network is independent of train type(class), it is conceived that the novel IoT sensor network is quickly deployed to various types of trains.

III. Conclusion

This brief paper proposes situation-aware collection and transmission of mission-critical sensor data to proactively avoid critical situation. Types of sensor data can be enumerated as tilting sensor, reader of position information on train, and various sensors in the interior of train vehicles. Depending on faced situation, data obtained from a sensor takes higher weight as compared to the data from other sensors. Also, the rate of transmission of sensor data is variably changed according to the situation. This methodology can be utilized for high speed trains. Particularly, a medium access control based on queue management, which is fit to sensor network for vehicles is proposed and performance evaluation is also performed.